# *Title page*

- **Title of the paper:**

  **PREFERENTIAL ENERGIZATION OF ALPHA PARTICLES IN POLAR CORONAL HOLES AT ONE SOLAR RADIUS ABOVE THE PHOTOSPHERE**

- **Names of the authors:**

    1. **ANIRUDDHA CHAKRAVARTY**
       **Department of Mathematics,**
       **Jadavpur University,**
       **Kolkata-700032,**
       **India.**

    2. **M. BOSE**
       **Department of Physics,**
       **Jadavpur University,**
       **Kolkata-700032,**
       **India.**

- **Corresponding author's *e-mail* address: mridulbose@gmail.com**

# Preferential energization of alpha particles in polar coronal holes at one solar radius above the photosphere

Aniruddha Chakravarty[1] and M. Bose[2*]

[1]*Department of Mathematics, Jadavpur University, Kolkata-700032, India*
[2]*Department of Physics, Jadavpur University, Kolkata-700032, India*

[*]Corresponding author's *e-mail* address: mridulbose@gmail.com

**ABSTRACT**
Heating of polar coronal holes (PCH) during solar minimum and acceleration of the fast solar wind issuing therefrom lack comprehensive theoretical understanding. Wave–particle interactions are considered to have crucial effects on the extreme properties of heavy ions in the collisionless region of the PCH. In this article, we have presented a novel sensitivity analysis to investigate plasma heating by radio waves at lower hybrid (LH) frequencies. We have employed a three-fluid Maxwell model comprising electrons, protons, and α-particles at around two solar radii heliocentric distance in the PCH and derived a dispersion relation as a 13th-order polynomial for the frequency. Our model provides indications of preferential heating of α-particles in comparison with protons by means of LH instabilities. We have employed the electron velocity and spatial charge distribution as our basic study tools so as to show the effects of alpha–proton differential mass and differential perpendicular velocity on the preferential heating of α-particles.

## 1 INTRODUCTION

Polar coronal holes (PCH) are regions of low-density plasma that persist for nearly 7 years around the minimum in the Sun's 11-year activity cycle (Cranmer 2009). The open magnetic flux tubes rooted in the PCH expand superradially (Cranmer 2001). The coronal magnetic field structure during solar minimum has been depicted by a semi-empirical dipole–quadrupole–current-sheet (DQCS) model (Banaszkiewicz, Axford & McKenzie 1998).

The identification of the PCH as the source of the fast solar wind (FSW) has been established convincingly. Based on *Ulysses* observations during solar minimum, it has been inferred that the FSW undergoes significant non-radial expansion (Marsden, Wenzel & Smith 1997). Zouganelis et al. (2005) have constructed a FSW speed profile as a function of the heliocentric distance by employing a kinetic model. In the FSW, coronal heating is proton and ion dominated (Hollweg 2008). *SOHO/UVCS* observations found strong temperature anisotropies of heavy ions in the PCH causing higher temperatures in directions perpendicular to the magnetic field than in parallel directions (Ofman 2004). In particular, *SOHO/UVCS* demonstrated preferential energization of the heavier OVI ions in comparison with the lighter HI atoms in the PCH (Kohl et al. 1998). In spite of the absence of *in-situ* observations of the coronal plasma, data from the *Helios* (Marsch et al. 1982) and *Messenger* (Gershman et al. 2012) spacecraft in the inner heliosphere have indicated that HeIII ions (α-particles) in the FSW are preferentially heated in relation to protons.

The extreme properties of heavy ions in the collisionless region of the PCH still lack satisfactory theoretical understanding. There are strong indications that, in order to maintain the observed plasma conditions in the FSW, an extended energy source is necessary to augment basal heating (Cranmer 2001), along with a momentum deposition mechanism operating in the PCH at heliocentric distances greater than 2 solar radii [$R_\odot$] (Zangrilli et al. 2002). It is generally believed that wave–particle interactions play crucial roles in the preferential perpendicular heating of heavy ions that ultimately drives the FSW. In a recent work, hereafter called paper 1, Chakravarty & Bose (2014) have investigated dissipation of lower hybrid (LH) waves via stochastic heating of α-particles at about $2R_\odot$ in the PCH. Isenberg & Vasquez (2007) proposed a kinetic mechanism, known as the Fermi mechanism, for the preferential perpendicular heating of minor ions in the PCH based on resonant cyclotron interaction between ions and parallel-propagating ion-cyclotron waves. Chandran et al. (2010) studied preferential stochastic heating of heavy ions by Alfvén wave/kinetic Alfvén wave turbulence in the solar wind.

In paper 1, we have obtained several wave number ranges corresponding to heating of α-particles in the PCH due to LH instabilities. This article is an extension of paper 1 with several refinements/modifications. As far as the new content of this article is concerned, here we have carried out a novel sensitivity analysis to investigate the preferential heating of α-particles over protons, due to LH instabilities, in relation to several input parameters. In order to have a more realistic situation, the restriction imposed on one parameter of our model in paper 1 has been slackened, and the resulting modified dispersion relation (DR) for plasma waves presented in Section 2. Comparative heating efficiency of protons and α-particles has been studied in Section 3 by using exact solutions of the DR. Discussion and conclusions are given in the last section.

## 2 WAVE DISPERSION RELATION

The PCH plasma consists mainly of electrons and protons. Protons and α-particles are considered to be the only major ion species, although the latter is much less abundant than protons (Isenberg & Vasquez 2007). The primary acceleration region of the plasma particles in the FSW starts approximately from $1.5R_\odot$ heliocentric distance (Cranmer 2002a). It is believed that collisionless wave mechanisms start to play a dominant role in the overall plasma dynamics at around $2R_\odot$ (Cranmer 2002b). Also, it is noted that at all heights in the corona, $\beta << 1$ (Cranmer & van Ballegooijen 2003), where $\beta$ denotes the plasma beta.

In order to investigate the preferential acceleration and heating of heavy ions, Hu, Esser & Habbal (2000) proposed a one-dimensional, four-fluid turbulence driven solar wind model composed of four species: electrons, protons, α-particles, and one minor ion species. Now, we note that in the low collisionality regime in a plasma with $\beta \ll 1$, fluid theory (MHD) may reasonably be employed to study the plasma dynamics, nearly perpendicular to the magnetic field, at large scales independent of the small ones (see Belmont 2008; Fitzpatrick 2011). As in paper 1, here we employ a three-dimensional, three-fluid Maxwell model comprising electrons, protons, and α-particles. In this article we investigate the preferential heating of α-particles over protons by means of LH instabilities. Our region of study at around $2R_\odot$ heliocentric distance in the PCH corresponds to the following assumptions that simplify our model. First, the ambient magnetic field has been taken along the z-direction in a cylindrical polar coordinate system; second, no momentum addition mechanism has been considered; third, same isotropic and uniform initial temperature has been assumed for each of the three particle species; and fourth, the basic flow velocity of each species is assumed to have the same initial direction.

The fluid equations (1) – (3) for the *s*th species are the equations of continuity, momentum, and state, given by

$$\frac{\partial n_s}{\partial t} + \nabla \cdot (n_s \boldsymbol{u_s}) = 0, \quad (1)$$

$$m_s n_s \left[ \frac{\partial \boldsymbol{u_s}}{\partial t} + (\boldsymbol{u_s} \cdot \nabla) \boldsymbol{u_s} \right]$$
$$= e_s n_s (\boldsymbol{E} + \boldsymbol{u_s} \times \boldsymbol{B}) - \nabla p_s, \quad (2)$$

and

$$p_s = n_s k_B T_s = n_s m_s V_{T_s}^2, \quad (3)$$

respectively, coupled with Maxwell's divergence equations and the curl $\boldsymbol{E}$ equation, where the subscript s = e, p, α stands for electrons, protons, and α-particles respectively. $e_s$, $m_s$, $n_s$, $p_s$, $V_{T_s}$, and $\boldsymbol{u_s}$ denote the charge (including sign), mass, number density, scalar pressure, thermal speed of particles with temperature $T_s$, and fluid velocity of the *s*th species, respectively. $\boldsymbol{E}$ is the electric field, $\boldsymbol{B}$ the magnetic flux density, and $k_B$ the Boltzmann constant. These six equations are the same as in paper 1.

Small first-order perturbations proportional to exp ($i\boldsymbol{k} \cdot \boldsymbol{r} - i\omega t$) are considered about an equilibrium, steady state, where $\boldsymbol{k}$ is the wavevector and $\omega$ the wave angular frequency. Uniform, constant equilibrium values of the dependent variables $n_s$, $\boldsymbol{u_s}$, $\boldsymbol{E}$, and $\boldsymbol{B}$ are $n_{s0}$ ($\tilde{n}_s$), $\boldsymbol{u_{s0}}$ ($\tilde{\boldsymbol{u}}_s$), $\boldsymbol{E_0}$ ($\tilde{\boldsymbol{E}}$), and $\boldsymbol{B_0}$ ($\tilde{\boldsymbol{B}}$), respectively, with the Fourier transforms of the corresponding perturbations given within parentheses.

The vector directional diagram in the cylindrical polar coordinate system is shown in Fig.1. $\boldsymbol{B_0}$ is in the *z*-direction and $\boldsymbol{E_0}$ is in the *r*–θ-plane. The average bulk outflow velocity due to the solar wind [$\boldsymbol{u_0}$] and $\boldsymbol{u_{s0}}$ have the same direction. The first point of difference from paper 1 is that $\boldsymbol{k}$ is taken in the *r*–*z*-plane here. The angular range of $\boldsymbol{k}$ relative to $\boldsymbol{B_0}$ is given quantitatively in the next section.

The *r*-component of the equation (2) yields, after much algebraic manipulation, the following DR:

$$\sum_s \frac{e_s^2 X(e_s Y - iZ)}{e_s^2 Y^2 + Z^2} = -i\varepsilon_0 \omega \frac{\cos\varphi_E \cos\eta_k}{\delta}, \quad (4)$$

where
$$X = n_{s0}[\omega \cos\varphi_E + k u_{s0}(\cos\psi_{u\theta} \sin\varphi_E \cos\eta_k$$
$$- \cos\psi_{uz} \cos\varphi_E \sin\eta_k)],$$

$$Y = k E_0 \delta \cos\varphi_E + \omega B_0 \cos\psi_{u\theta},$$

$$Z = m_s [(\omega - k\, u_{s0}\, \delta\,)^2 \cos\psi_{ur} - (kV_{Ts})^2\, \delta \cos\eta_k],$$

$$\delta = \cos\psi_{ur} \cos\eta_k + \cos\psi_{uz} \sin\eta_k,$$

and

$$u_{s0} = u_{s0\perp}\, (u_0/u_{0\perp}),$$

$$\cos\psi_{u\theta} = \pm \sqrt{(u_{0\perp}/u_0)^2 - \cos^2\psi_{ur}},$$

$$\cos\psi_{uz} = \pm \sqrt{1 - (u_{0\perp}/u_0)^2}.$$

$e_e = -e$, $e_p = +e$, and $e_\alpha = +2e$. $\varepsilon_0$ is the permittivity of free space, and $k$ the wave number. $\psi_{ur}$, $\psi_{u\theta}$, and $\psi_{uz}$ are the direction angles of the line along $u_0$, i.e. along $u_{s0}$ as well. $\varphi_E$ and $\eta_k$ are the angles made with the $r$-direction by $E_0$ and $k$ respectively. The subscript $\perp$ indicates a direction perpendicular to $B_0$.

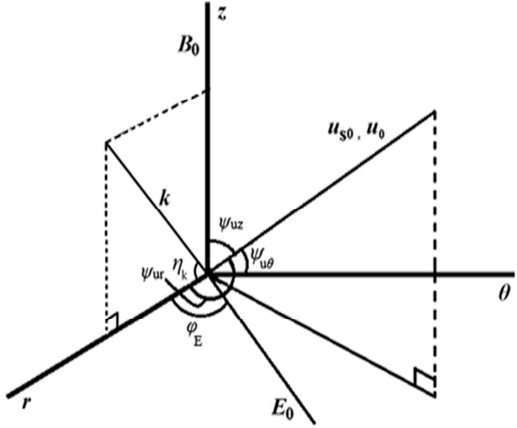

**Figure 1.** Vector directional diagram in cylindrical polar coordinates with $r,\theta,z$ as the orthonormal basis vectors.

The DR (4) is a 13th-order polynomial for $\omega$ as a function of real $k$ and it is a modified form of the DR presented in paper 1.

## 3 SENSITIVITY ANALYSIS

In order to investigate the comparative perpendicular heating efficiency of protons and α-particles, we model the time-averaged structure and composition of the PCH during solar minimum at about $2R_\odot$ heliocentric distance. First, relevant basic input parameters, subject to the current observational uncertainties, are listed. The values of several other necessary parameters are then derived with the help of these basic parameters.

*Basic Parameters*:

We note that the values of $T_s$, $B_0$, $n_{e0}$, $\varphi_E$, $u_0$, and $u_{p0\perp}$, given below, have been taken from paper 1.
(i) $T_s \approx 10^6$ K (s = e, p, α)
(ii) $B_0 \approx 10^{-4}$ T
(iii) $n_{e0} \approx 5\times 10^{11}$ m$^{-3}$.

The ratio of the number density of α-particles to protons remains fairly steady at about 0.048 in PCH flows (Neugebauer 1993; Laming 2004). We used an LH resonance formula in paper 1 to obtain $n_{\alpha 0}$ as about 1.62 per cent of $n_{p0}$. That formula has been omitted in this article and the abundance of α-particles has been taken as about 4.8 per cent of $n_{p0}$. Thus modified values of $n_{p0}$ and $n_{\alpha 0}$ are obtained as $4.56 \times 10^{11}$ m$^{-3}$ and $2.19 \times 10^{10}$ m$^{-3}$ respectively.
(iv) $\varphi_E \approx \pi/4$. $\psi_{ur}$, $\psi_{u\theta}$, and $\psi_{uz}$ are positive acute angles as in paper 1.
(v) $u_0 \approx 1.60 \times 10^5$ m s$^{-1}$.

The α-particles flow faster than the protons at around $2R_\odot$ heliocentric distance and there the perpendicular velocities of protons and O VI ions are about 200 and 300–400 km s$^{-1}$ respectively (see Laming & Lepri 2007, and references therein). *Helios* observations indicate that the alpha–proton differential speed is as high as 150 km s$^{-1}$ near the Sun (Reisenfeld et al. 2001). We have taken $u_{p0\perp} \approx 2\times 10^5$ m s$^{-1}$. Here we have assumed three values of $u_{\alpha 0\perp}$ [in m s$^{-1}$], given by $4 \times 10^5$, $2.5 \times 10^5$, and $2 \times 10^5$, while only the foremost value was considered in paper 1. We define the differential perpendicular velocity (DPV) between protons and α-particles as

$$\Delta u_\perp = u_{\alpha 0\perp} - u_{p0\perp}.$$

Thus apart from the zero value, two positive values of $\Delta u_\perp$, $2\times 10^5$ m s$^{-1}$ and $5\times 10^4$ m s$^{-1}$, have been considered here. Only two values of $u_{e0\perp}$, $5 \times 10^3$ m s$^{-1}$ and $5 \times 10^4$ m s$^{-1}$, were assumed in paper 1. Instead, we have considered here a wide range of $u_{e0\perp}$: $5 \times 10^3$–$4 \times 10^4$ m s$^{-1}$ for our analysis.

*Derived Parameters*:

(i) We have

$$V_{Ts} = \sqrt{k_B T_s / m_s}$$

so that $V_{Te} \approx 3.89 \times 10^6$ m s$^{-1}$, $V_{Tp} \approx 9.09 \times 10^4$ m s$^{-1}$, and $V_{T\alpha} \approx 4.55 \times 10^4$ m s$^{-1}$.

(ii) Particle gyrofrequency and particle plasma frequency are given [in rad s$^{-1}$] by

$$\Omega_{cs} = \frac{|e_s| B_0}{m_s} \quad \text{and} \quad \omega_{ps} = \sqrt{\frac{n_{s0} e_s^2}{\varepsilon_0 m_s}}$$

respectively, with $B_0$ expressed in tesla (see paper 1). Then we obtain [in rad s$^{-1}$]

$\Omega_{ce} \approx 1.76 \times 10^7$, $\Omega_{cp} \approx 9.60 \times 10^3$, $\Omega_{c\alpha} \approx 4.79 \times 10^3$, $\omega_{pe} \approx 4.00 \times 10^7$, $\omega_{pp} \approx 8.90\times 10^5$, and $\omega_{p\alpha} \approx 1.17\times 10^5$.

(iii) The LH range of frequencies [$\omega_{LH,i}$] satisfies (Bonoli 1984)

$$\Omega_{ci}^2 \ll \omega_{LH,i}^2 \ll \Omega_{ce}^2$$

and

$$\omega_{LH,i} = \sqrt{\Omega_{ce}\Omega_{ci}}, \quad (i = p,\alpha)$$

so that
$\omega_{LH,p} \approx 4.11\times 10^5$ rad s$^{-1}$ and
$\omega_{LH,\alpha} \approx 2.90\times 10^5$ rad s$^{-1}$.

(iv) As a refinement of the model in paper 1, here we consider that LH waves propagate across the magnetic field lines satisfying the condition (Laming 2004)

$$\sin\eta_k = \frac{k_\parallel}{k} < \frac{\omega_{pi}}{\omega_{pe}}, \quad (i = p,\alpha)$$

where $k_\parallel$ is the component of $\boldsymbol{k}$ parallel to $\boldsymbol{B_0}$. For proton–electron combination, we obtain $\sin\eta_k < 2.23 \times 10^{-2}$, and so we may take $\sin\eta_k \approx 2 \times 10^{-2}$ and $\cos\eta_k \approx 1$. Again, for alpha–electron combination, $\sin\eta_k < 2.93 \times 10^{-3}$, and in this case we take $\sin\eta_k \approx 2\times 10^{-3}$ and $\cos\eta_k \approx 1$.

(v) The plasma beta is given by

$$\beta = (2\mu_0 / B_0^2)\left(\sum_s n_{s0} k_B T_s\right),$$

where $\mu_0$ is the permeability of free space. We note that $\beta \sim 3.14\times 10^{-3}$ at the base of the collisionless region of the PCH (Isenberg & Vasquez 2007). In our case, we have $\beta \approx 3.39\times 10^{-3}$.

The heating of α-particles and protons is now studied separately under the following two subsections, where the effects of the DPV's as well as the differential mass (DM) between these two ion species are taken into account.

### 3.1 Heating of α-particles

In order to study the heating of α-particles, we have considered the following three cases:

**Case I. Effects of DM and high DPV.**

In this case, $u_{p0\perp} \approx 2 \times 10^5$ m s$^{-1}$ and $u_{\alpha 0\perp} \approx 4 \times 10^5$ m s$^{-1}$ give $\Delta u_\perp \approx 2 \times 10^5$ m s$^{-1}$.

Taking $u_{0\perp}$ as the weighted average of $u_{s0\perp}$ with $n_{s0}$ as weights, we find $u_{0\perp}/u_0 \approx 0.65$–$0.77$, where $u_{e0\perp} \approx 5 \times 10^3 - 4 \times 10^4$ m s$^{-1}$. We take $\cos\psi_{ur} \approx 0.50 < u_{0\perp}/u_0$ (see Section 2).

Particle gyroradius is given by

$$r_{cs} = \frac{u_{s0\perp}}{\Omega_{cs}}. \quad (s = e, p, \alpha)$$

Then we obtain
$r_{ce} \approx 2.84 \times 10^{-4} - 2.27 \times 10^{-3}$ m and
$r_{c\alpha} \approx 8.35 \times 10$ m
giving the range of the cyclotron resonance wave number for electrons: $2.77 \times 10^3 - 2.21 \times 10^4$ m$^{-1}$ and the cyclotron resonance wave number for α-particles $[k_{IC,\alpha}]$ as $7.52 \times 10^{-2}$ m$^{-1}$. We consider the range of LH wave number for α-particles $[k_{LH,\alpha}]$ as $0.8$–$3.0$ m$^{-1}$.

Before showing that energy redistribution in the plasma due to LH instabilities favors α-particles over protons, we need to clarify the criteria adopted in our analysis that must be satisfied for heating of the concerned ion species.

First, we define the strength of instability of standing waves as (see paper 1)

$$\frac{\tau}{P} = \frac{1}{2\pi}\left|\frac{\omega_r}{\omega_i}\right|,$$

where $\omega_r$ and $\omega_i$ are the real and imaginary parts of $\omega$ respectively. The wave period is $P = 2\pi / |\omega_r|$ and the decay (or growth) time of wave is $\tau = 1/|\omega_i|$. It is clear that stronger wave instability corresponds to lower value of $\tau/P$ and vice versa.

We note that the sufficient condition for significant net heating (SNH) of α-particles is the occurrence of any one of the following two combinations of LH wave modes in a particular solution set of DR (4).

**Combination (i).** When there is only one LH wave mode which is a decaying mode, satisfying both the following two criteria.

(a) Criterion of resonant LH frequency:
Resonant LH waves are taken as those with $2.4 \times 10^5 \leq \omega_r \leq 3.4 \times 10^5$ rad s$^{-1}$ for α-particles, where it may be recalled that $\omega_{LH,\alpha} \approx 2.90 \times 10^5$ rad s$^{-1}$.

(b) Criterion of instability strength:
$$0.80 \leq \tau/P \leq 1.90$$

**Combination (ii).** When there is one decaying LH mode along with one growing mode, satisfying the criterion (c) given below, in addition to the above two criteria (a) and (b).

(c) Criterion of differential instability strength:
$$\tau_g/P - \tau_d/P \geq 0.01,$$
where $\tau_d/P$ and $\tau_g/P$ are the strengths of decay and growth of waves respectively.

In order to investigate the preferential LH heating of α-particles in comparison with protons in an analytical way, we have employed MATHEMATICA 4.1 (Wolfram 2003) for solving the DR (4).

*Algorithm for heating of α-particles:*

For each value of $u_{e0\perp}$, $E_0$ is increased from 10 V m$^{-1}$ in steps of 0.5 V m$^{-1}$. For each such $E_0$, we seek one value of $k$ by increasing it in steps of 0.01 m$^{-1}$ over its entire range of 0.8–3.0 m$^{-1}$, so that the DR (4) yields SNH of α-particles. If there is no such value of $k$, we take the next higher value of $E_0$ and again we seek one value of $k$ corresponding SNH. These steps are repeatedly followed. Now, the value of $E_0$ at one step-size less than that of the earliest occurrence of SNH, is termed as the critical electric field for heating of α-particles [$E_{0c,\alpha}$]. $u_{e0\perp}$ is increased in its range 5–40 × $10^3$ m s$^{-1}$, in steps of 2.5 × $10^3$ m s$^{-1}$, and $E_{0c,\alpha}$ is obtained corresponding to each $u_{e0\perp}$. $E_{0c,\alpha}$ may be taken as a measure of the heating efficiency of α-particles. It is clear that lower the value of $E_{0c,\alpha}$, higher is the heating efficiency of α-particles and vice versa. The points ($u_{e0\perp}$, $E_{0c,\alpha}$) are plotted and a quartic polynomial in $u_{e0\perp}$ fits to the data as shown by a dot-dashed line with open triangles in Fig. 2. This is the regression line representing the heating-efficiency profile (HEP) for α-particles.

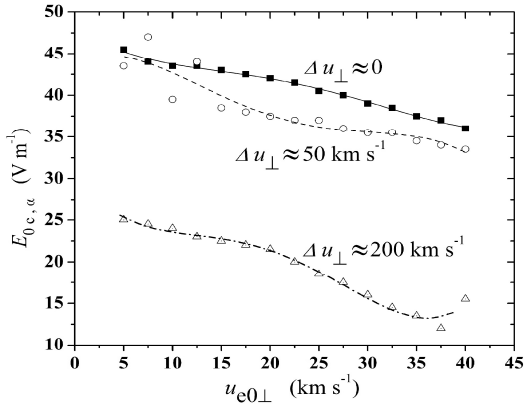

**Figure 2.** Heating efficiency profiles for α-particles.

In fact, SNH of α-particles has been found to occur for each $E_0$ corresponding to a particular $u_{e0\perp}$, where $E_0$ is increased, in steps of 0.5 V m$^{-1}$, from ($E_{0c,\alpha}$ + 0.5) V m$^{-1}$ to 80 V m$^{-1}$. This is true for each $u_{e0\perp}$ as its value is increased, in steps of 2.5 × $10^3$ m s$^{-1}$, from 5 × $10^3$ m s$^{-1}$ to 3.5 × $10^4$ m s$^{-1}$. For $u_{e0\perp} \approx 3.75 \times 10^4$ m s$^{-1}$, $E_{0c,\alpha} \approx 12$ V m$^{-1}$; it is noted that SNH occurs for each $E_0$ in its range 12.5–80 V m$^{-1}$, where $E_0$ is increased in steps of 0.5 V m$^{-1}$, except for values of $E_0$ [in V m$^{-1}$]: 13–13.5, 14.5, 15.5–16, 17, and 18–18.5. For $u_{e0\perp} \approx 4 \times 10^4$ m s$^{-1}$, $E_{0c,\alpha} \approx 15.5$ V m$^{-1}$; SNH occurs for each $E_0$ in the range 16–80 V m$^{-1}$, where $E_0$ is increased in steps of 0.5 V m$^{-1}$, except for values of $E_0$ : 16.5–17.5 V m$^{-1}$. We ignore these irregularities/anomalies, as these cases constitute a small percentage of the total number of cases considered, and so may be taken to be within the limits of methodological uncertainties involved in obtaining the HEP.

The zone above the HEP, corresponding to $E_{0c,\alpha} < E_0 \leq 80$ V m$^{-1}$, may be considered as the heating zone for α-particles in this case.

**Case II. Effects of DM and low DPV.**

In this case, $\Delta u_\perp \approx 5 \times 10^4$ m s$^{-1}$, where $u_{\alpha 0\perp} \approx 2.5 \times 10^5$ m s$^{-1}$. Also, $u_{0\perp}/u_0 \approx 0.63$–0.75, $r_{c\alpha} \approx 5.22 \times 10$ m, $k_{IC,\alpha} \approx 1.20 \times 10^{-1}$ m$^{-1}$, and the range of $k_{LH,\alpha}$: 1.45–4.00 m$^{-1}$.

Other relevant parameters/considerations are the same as in case I. The HEP is shown as a dashed line with open circles in Fig. 2.

SNH for α-particles occurs for each $E_0$ corresponding to a particular $u_{e0\perp}$, as $E_0$ is increased from ($E_{0c,\alpha}$ + 0.5) V m$^{-1}$ to 80 V m$^{-1}$. This is true for each $u_{e0\perp}$, as we increase $u_{e0\perp}$ from 7.5 × $10^3$ m s$^{-1}$ to 4 × $10^4$ m s$^{-1}$. For $u_{e0\perp} \approx 5 \times 10^3$ m s$^{-1}$, $E_{0c,\alpha} \approx 43.5$ V m$^{-1}$; SNH is found to occur for each $E_0$ in the range 44–80 V m$^{-1}$ except for values of $E_0$ : 47–48.5 V m$^{-1}$.

**Case III. Effects of DM only.**

In this case, $u_{\alpha 0\perp} \approx 2\times 10^5$ m s$^{-1}$ so that $\Delta u_\perp \approx 0$. Also, $u_{0\perp}/u_0 \approx 0.63$–0.74, $r_{c\alpha} \approx 4.18\times 10$ m, $k_{IC,\alpha} \approx 1.50\times 10^{-1}$ m$^{-1}$, and the range of $k_{LH,\alpha}$: 1.70–4.20 m$^{-1}$.

The HEP is shown as a solid line with filled squares in Fig. 2.

SNH of α-particles occurs for each $E_0$ corresponding a particular $u_{e0\perp}$, where $E_0$ is increased from ($E_{0c,\alpha}$ + 0.5) V m$^{-1}$ to 80 V m$^{-1}$. This is true for each value of $u_{e0\perp}$, as it is increased from $5 \times 10^3$ m s$^{-1}$ to $4 \times 10^4$ m s$^{-1}$.

## 3.2 Heating of protons

Once again, the same three cases as in section 3.1 are considered.

**Case I. Effects of DM and high DPV.**

Only the parameter values differing from those given under case I for α-particles in section 3.1 are listed here.

$r_{cp} \approx 2.08\times 10$ m$^{-1}$ gives the cyclotron resonance wave number for protons as $3.02\times 10^{-1}$ m$^{-1}$. We consider the LH wave number range for protons [$k_{LH,p}$] as 3.2–6.0 m$^{-1}$. Resonant LH waves are taken as those with $3.6\times 10^5 \leq \omega_r \leq 4.6\times 10^5$ rad s$^{-1}$ for protons, where we have $\omega_{LH,p} \approx 4.11\times 10^5$ rad s$^{-1}$. We take the starting value of $E_0$ as 50 V m$^{-1}$. We employ the same algorithm used for investigating the heating of α-particles as given in section 3.1 and plotted the points ($u_{e0\perp}$, $E_{0c,p}$) in order to obtain the HEP for protons as shown by a dot-dashed line with open triangles in Fig. 3.

SNH of protons occurs for each $E_0$ corresponding a particular $u_{e0\perp}$, where $E_0$ is increased from ($E_{0c,p}$ + 0.5) V m$^{-1}$ to 80 V m$^{-1}$. This is true for each $u_{e0\perp}$, as its value is increased from $5 \times 10^3$ m s$^{-1}$ to $4 \times 10^4$ m s$^{-1}$.

The HEP for protons in case II, regarding the combined effects of DM and low DPV, is shown as a dashed line with open circles in Fig. 3.

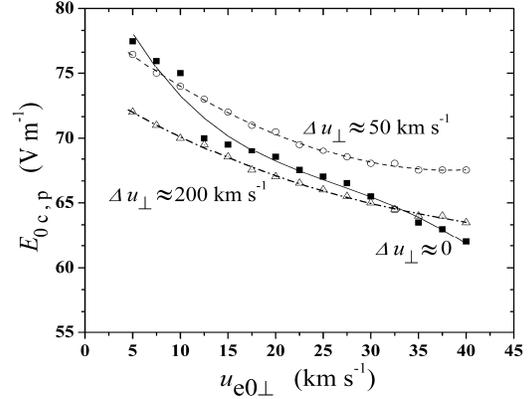

**Figure 3.** Heating efficiency profiles for protons.

The solid line with filled squares in Fig. 3 shows the HEP for protons in case III, regarding the effects of DM only.

With the help of the HEP's in Fig. 2 and Fig. 3, preferential heating of α-particles over protons has been presented in the following section.

## 4 DISCUSSION AND CONCLUSIONS

In fusion devices, heating by LH instabilities is considered to be a promising method of plasma heating through ion heating in the radio frequency range (Bose 1995, and references therein). In an astrophysical situation, it has been argued that Alfvén waves having frequency less than ion-cyclotron frequency may directly undergo polarization drift so as to generate high frequency LH waves, and these driven LH waves may be effective in preferential transverse heating and acceleration of ions in PCH (Singh & Khazanov 2004).

We now discuss various results of our analysis, where the LH heating regimes for α-particles and protons are 38.2–54.1 kHz and 57.3–73.2 kHz respectively, as obtained from the ranges of $\omega_r$ for these two ion species (see case I under sections 3.1 and 3.2).

Each curve in Fig. 2 depicts the respective threshold conditions for LH instabilities so that the zone above each curve, where $E_0 \leq 80$ V m$^{-1}$, shows the heating zone for α-particles in relation to $u_{e0\perp}$. These curves are termed as HEP's for α-particles. Similar HEP's for protons are shown in Fig. 3, where $E_{0c,p} < E_0 \leq 80$ V m$^{-1}$ signifies the heating zones for protons under the corresponding conditions.

The essence of our sensitivity analysis is to study, in a comparative manner, the LH heating of protons and α-particles under different conditions, with the aid of the functional relationship between two parameters, $u_{e0\perp}$ and $E_{0c,i}$ [i = p,α]. The following important points may be noted from Fig. 2 and Fig. 3.

(i) The HEP for α-particles for $\Delta u_\perp \approx 2 \times 10^5$ m s$^{-1}$ lies below that for $\Delta u_\perp \approx 5 \times 10^4$ m s$^{-1}$ for the entire range of $u_{e0\perp}$ (see Fig. 2). The HEP's for $\Delta u_\perp \approx 0$ and $5 \times 10^4$ m s$^{-1}$ almost coincide for low values, below about $10^4$ m s$^{-1}$, of $u_{e0\perp}$. Thereafter, for the remaining ranges of $u_{e0\perp}$, the HEP for $\Delta u_\perp \approx 5 \times 10^4$ m s$^{-1}$ lies below that for $\Delta u_\perp \approx 0$. This implies that an increase in the value of DPV from zero causes enhancement of the heating efficiency of α-particles over almost the entire range of $u_{e0\perp}$ considered for our analysis.

(ii) The HEP for protons for $\Delta u_\perp \approx 2 \times 10^5$ m s$^{-1}$ lies below that for $\Delta u_\perp \approx 5 \times 10^4$ m s$^{-1}$ for the entire range of $u_{e0\perp}$ (see Fig. 3). It is also found that the slope of both these HEP's tends from negative to zero as $u_{e0\perp}$ increases. There is a lateral separation between these two HEP's over the entire range of $u_{e0\perp}$. In a sharp contrast, the HEP for $\Delta u_\perp \approx 0$, although starts above the HEP for $\Delta u_\perp \approx 5 \times 10^4$ m s$^{-1}$, goes downward across both the HEP's for positive DPV's, as $u_{e0\perp}$ increases. The HEP for $\Delta u_\perp \approx 0$ intersects the HEP's for $\Delta u_\perp \approx 5 \times 10^4$ m s$^{-1}$ and $2 \times 10^5$ m s$^{-1}$ at $u_{e0\perp} \approx 7.5 \times 10^3$ and $3.3 \times 10^4$ m s$^{-1}$ respectively and remains below the HEP for $\Delta u_\perp \approx 2 \times 10^5$ m s$^{-1}$. These results imply that in the absence of DPV, DM exhibits its effects in the unhindered increase in the heating efficiency of protons with increase in $u_{e0\perp}$. Once positive DPV sets in, the dependence of heating efficiency of protons on $u_{e0\perp}$ decreases with increasing $u_{e0\perp}$, although the heating efficiency of protons is greater for the higher positive DPV than that for the lower positive DPV over the entire range of $u_{e0\perp}$.

Now, we define the alpha–proton differential critical electric field [$\Delta E_c$] for particular sets of $\Delta u_\perp$ and $u_{e0\perp}$ as

$$\Delta E_c = E_{0c,p} - E_{0c,\alpha},$$

where $E_{0c,\alpha}$ and $E_{0c,p}$ are obtained from Fig. 2 and Fig 3 respectively. We note that lower the value of $E_{0c,i}$, where i = p, α, higher is the heating efficiency of the concerned ion species i and vice versa. Again, positive value of $\Delta E_c$ implies preferential heating of α-particles in comparison with protons for particular sets of $\Delta u_\perp$ and $u_{e0\perp}$. Clearly, the preferential heating gets stronger as the positive value of $\Delta E_c$ increases.

It is easily seen from the two dot-dashed curves in Fig. 2 and Fig.3, corresponding to $\Delta u_\perp \approx 2 \times 10^5$ m s$^{-1}$, that $\Delta E_c$ is always positive for the entire range of $u_{e0\perp}$: $5$–$40 \times 10^3$ m s$^{-1}$. Similar comparison in these figures between the two dashed curves corresponding to $\Delta u_\perp \approx 5 \times 10^4$ m s$^{-1}$ as well as between the two solid curves corresponding to $\Delta u_\perp \approx 0$ always yields positive values of $\Delta E_c$. This indicates that α-particles are preferentially heated in comparison with protons for each value of $\Delta u_\perp$ and for the entire range of $u_{e0\perp}$ considered for our analysis.

Now, we measure the strength of preferential heating of α-particles as a percentage of the heating efficiency of protons for particular sets of $\Delta u_\perp$ and $u_{e0\perp}$ by using the quantity

$$H_\alpha = (\Delta E_c / E_{0c,p}) \times 100 \ .$$

For $\Delta u_\perp \approx 2 \times 10^5$ m s$^{-1}$, the points ($u_{e0\perp}$, $H_\alpha$) are plotted and a quartic polynomial in $u_{e0\perp}$ fits to the data as shown by a dot-dashed line with open triangles in Fig. 4. This regression line represents the functional relationship of the strength of preferential heating of α-particles over protons with $u_{e0\perp}$. Similar curves for $\Delta u_\perp \approx 5 \times 10^4$ m s$^{-1}$ and $\Delta u_\perp \approx 0$ are shown by a dashed line with open circles and a solid line with filled squares respectively in Fig. 4.

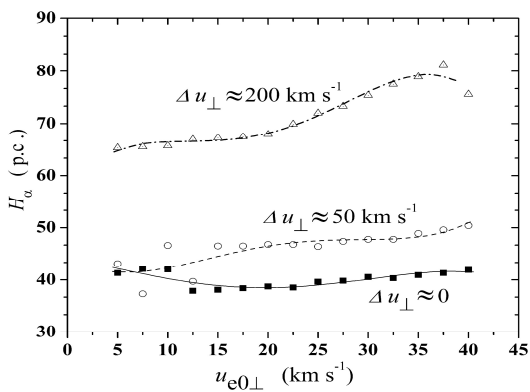

**Figure 4.** Preferential heating of α-particles over protons for different values of $\Delta u_\perp$.

It is observed from Fig. 4 that The curve for $\Delta u_\perp \approx 5 \times 10^4$ m s$^{-1}$ lies below that for $\Delta u_\perp \approx 2 \times 10^5$ m s$^{-1}$ for the entire range of $u_{e0\perp}$. Except for low values of $u_{e0\perp}$, below about $10^4$ m s$^{-1}$, the curve for $\Delta u_\perp \approx 0$ lies below that for $\Delta u_\perp \approx 5 \times 10^4$ m s$^{-1}$ for the remaining range of $u_{e0\perp}$. This implies that for a particular value of $u_{e0\perp}$ above $\approx 10^4$ m s$^{-1}$, the strength of preferential heating of α-particles over protons increases with DPV.

Interesting outcomes of our analysis are summarized here.

(i) In the absence of the alpha–proton DPV, DM affects the system in such a way that the heating efficiency of protons goes on increasing with $u_{e0\perp}$, and eventually becomes greater than that for the high DPV of $2 \times 10^5$ m s$^{-1}$ near the upper end of the range of $u_{e0\perp}$ (see Fig. 3). In contrast, the absence of DPV depicts that the heating efficiency of α-particles exhibits restricted increase and remains less than that for the low DPV of $5 \times 10^4$ m s$^{-1}$ over almost the entire range of $u_{e0\perp}$ (see Fig. 2). Still we find that DM alone can result in preferential heating of α-particles over protons for the entire range of $u_{e0\perp}$ (see the solid curve in Fig. 4).

(ii) Once positive alpha–proton DPV sets in, i.e. α-particles becomes faster than protons, the dependence of heating efficiency of protons on $u_{e0\perp}$ decreases with increasing $u_{e0\perp}$ (see Fig. 3). The heating efficiency of both these ion species is greater for the higher positive DPV than that for the lower positive DPV for the entire range of $u_{e0\perp}$ (see Fig. 2 and Fig. 3). Also, the preferential heating of α-particles over protons is stronger for the higher positive DPV than that for the lower positive DPV for the entire range of $u_{e0\perp}$ (see Fig. 4).

With a view to have a better understanding of the observed alpha–proton differential streaming in the FSW, our model attempts to investigate the preferential heating of α-particles in comparison with protons due to perpendicular magnetic anisotropy, where electron velocity and spatial charge distribution have been employed as our basic study tools. Although we have neglected the cross-scale phenomena like turbulence, our model, assuming fluid-approximation in a nearly collisionless plasma medium at about 1$R_\odot$ above the photosphere in the PCH, suggests that both of the alpha–proton DM and DPV contribute to the preferential heating of α-particles in comparison with protons by means of LH instabilities in the system. This result may have important bearing on the evolution of the non-linear collisionless plasma dynamics in the FSW.

The computational model proposed in this research problem does not include any minor ion species present in the PCH. As measured remotely by *SOHO*/*UVCS*, the primary observable minor ion species in the collisionless PCH is OVI (Isenberg &Vasquez 2007). It would be interesting to include OVI in our model to study its dispersive effects on α-particles and on itself too. This will be taken up as our next work.

*In-situ* measurements corresponding to FSW are available to some extent, but that too at distances not less than about 0.3au from the Sun. So the current knowledge about the crucial aspects of the plasma energetics and kinematics in the extended corona in FSW streams happens to be only in piecemeal. This inevitably involves a degree of uncertainty in our analytical results. *Solar Probe Plus*, an upcoming solar orbiter mission with periapsis at about 9.5$R_\odot$ heliocentric distance, will hopefully usher a new era by providing much more realistic constraints for the improvement of theoretical models.

## ACKNOWLEDGEMENTS

AC is thankful to ex-Prof. Dr. R. Bondyopadhaya of J.U. for valuable and inspiring discussions.

## REFERENCES

Banaszkiewicz M., Axford W.I., McKenzie J.F., 1998, A & A, 337, 940

Belmont G., 2008, in Collisionless Phenomena in Space Plasma Physics. amrel.obspm.fr/~ciardi/xla/abstracts/belmont.pdf

Bonoli P., 1984, in Linear Theory of Lower Hybrid Heating. MIT Report, PFC/RR-84-5

Bose M., 1995, Plasma Phys. Control. Fusion, 37, 223

Chakravarty A., Bose M., 2014, Sol. Phys., 289, 911

Chandran B.D.G., Li B., Rogers B.N., Quataert E., Germaschewski K., 2010, ApJ, 720, 503

Cranmer S.R., 2001, in Murdin, P., ed., Encyclopedia of Astron. and Astrophys. Nature Publishing Group/IOP Publishing, London/Bristol, article 1999. doi: 10.1888/0333750888/1999

Cranmer S.R., 2002a, Space Sci. Rev., 101, 229

Cranmer S.R., 2002b, in Wilson A., ed., From Solar Minimum to Solar Maximum. SP-508, ESA, 361

Cranmer S.R., 2009, Living Rev. Sol. Phys. 6(3). doi: 10.12942 / lrsp-2009-3

Cranmer S.R., van Ballegooijen A.A., 2003, ApJ, 594, 573

Fitzpatrick R., 2011, in Closure in Collisionless Magnetized Plasmas. http://farside.ph.utexas.edu/teaching/plasma/lectures/node40.html

Gershman D.J. et al., 2012, JGR, 117, A00M02

Hollweg J.V., 2008, J. Astrophys. Astron., 29, 217

Hu Y.Q., Esser R., Habbal S.R., 2000, JGR, 105, A3, 5093

Isenberg P.A., Vasquez B.J., 2007, ApJ, 668, 546

Kohl L. et al., 1998, ApJ, 501, L127

Laming J.M., 2004, ApJ, 604, 874

Laming J.M., Lepri S.T., 2007, ApJ, 660, 1642

Marsch E., Mühlhäuser K.-H., Rosenbauer H., Schwenn R., Neubauer F.M., 1982, JGR, 87, 35

Marsden R.G., Wenzel K.-P., Smith E.J., 1997, ESA Bull. 92, 75

Neugebauer M., 1993, in Observations of the Solar Wind from Coronal Holes. http://hdl.handle.net/2014/36520

Ofman L., 2004, in Lacoste H., ed., Proc. SOHO 13, Waves, Oscillations and Small-Scale Transient Events in the Solar Atmosphere : A Joint View from SOHO and Trace. SP-547, ESA, 345

Reisenfeld D.B., Gary S.P., Gosling J.T., Steinberg J.T., McComas D.J., Goldstein B.E., Neugebauer M., 2001, JGR, 106, 5693

Singh N., Khazanov G., 2004, JGR, 109, A05210. doi : 10.1029/2003JA010251

Wolfram S., 2003, The Mathematica Book, 5th ed., Wolfram media

Zangrilli L., Poletto G., Nicolosi P., Noci G., Romoli M., 2002, ApJ, 574, 477

Zouganelis I., Meyer-Vernet N., Landi S., Maksimovic M., Pantellini F., 2005, ApJ, 626, L117